\documentstyle[12pt, a4wide]{article}

\def\bea{\begin{eqnarray}}
\def\eea{\end{eqnarray}}
\def\beann{\begin{eqnarray*}}
\def\eeann{\end{eqnarray*}}
\def\beq{\begin{equation}}
\def\eeq{\end{equation}}
\def\ba{\begin{array}}
\def\ea{\end{array}}
\def\ben{\begin{enumerate}}
\def\een{\end{enumerate}}

\font\mybb=msbm10 at 11pt

\def\bb#1{\hbox{\mybb#1}}

\def\bR {\bb{R}}

\date{}

\begin{document}

\thispagestyle{empty}
\rightline{CERN-TH/2000-280}
\begin{center}
\vspace {.7cm}
{\large {\bf{ A no-go theorem for string warped compactifications}} }
\vskip  1.0truecm
{\large{\bf{ S. Ivanov${}^1$ and G. Papadopoulos${}^2$}}}
\vskip 2.0truecm
 {\normalsize{\sl 1. Department of Mathematics}}
\\ {\normalsize{\sl University of Sofia}}
\\ {\normalsize{\sl \lq\lq St. Kl. Ohridski''}}
\vskip 1.0truecm
 {\normalsize{\sl 2. CERN}}
\\ {\normalsize{\sl Theory Division}}
\\ {\normalsize{\sl 1211 Geneva, 23}}
\\ {\normalsize{\sl Switzerland}}

\vskip 2.0truecm
{\bf {Abstract}}
\end{center}
We give necessary conditions for the existence 
of perturbative  heterotic
and common sector type II string warped
 compactifications preserving four and
eight supersymmetries to four spacetime 
dimensions, respectively. In particular,
we find that the only compactifications of heterotic 
string with the spin connection
embedded in the gauge connection and type II strings are those
on Calabi-Yau manifolds with constant dilaton. 
We obtain similar results for compactifications to six and to two
dimensions.
\newpage

In past few years there has been interest in investigating
perturbative \cite{strom} and non-perturbative 
string warped compactifications
\cite{becker, vafa, gukov, sethi}.
One of the attractive features of such compactifications
is the  appearance of a scalar potential in the effective
low dimensional effective action  depending on 
some of moduli fields.  This  leads to the lifting of  some 
of the flat directions of the associated compactifications without
a warp factor.
The presence of the potential can be understood in various ways.
One way is that in  warped compactifications 
some ten- or eleven-dimensional supergravity
form field strengths receive an expectation value. Therefore
one can view the presence of the potentials as a consequence
of a Scherk-Schwarz type of mechanism.
Another feature of these warped compactifications is that
they preserve a fraction of spacetime supersymmetry.
Warped compactifications
have also  been related to Randall-Sundrum type of senario \cite{ver}.

Warped compactifications of the perturbative heterotic string
and of the common sector of type II  strings
have been investigated sometime ago by Strominger in \cite{strom}.
This was achieved by allowing the dilaton to be a non-constant
function of the compact space.
Necessary conditions were given for the existence
of such compactifications. In particular it was shown that
the compact manifold  of  such warped
 compactification is  a 2n-dimensional
 hermitian, but no K\"ahler, manifold equipped with a holomorphic
(n, 0)-form. Therefore it is required that  the 
Hodge number  $h^{n,0}={\rm dim}H^{n,0}$ of the compact manifold
is $h^{n,0}\geq 1$.

In this paper, we shall investigate the conditions
for the existence of holomorphic $(n,0)$-forms
on hermitian manifolds using the 
Gauduchon theorem \cite{G3}-\cite{G4}.
We shall assume  the following:

\begin{itemize}
\item[$\bullet$] The solution associated with the warped 
compactification is a smooth and the internal manifold
is compact.

\item[$\bullet$] For compactifications to 
four-dimensions the solution
preserves four and eight supersymmetries 
for the heterotic and type II
strings, respectively. The non-vanishing 
fields are those of the
common sector.

\item[$\bullet$] The dilaton is a globally 
defined scalar function on
the compact space.
\end{itemize}
Then we shall show that there are restrictions for
the existence of such forms which manifest 
themselves as conditions on
the string three-form field strength $H$ and its 
exterior derivative $dH$. 
In particular, we shall find that heterotic warped
compactifications with the spin connection
embedded in the gauge connection and type II string warped
compactifications are ruled out.
So the only compactifications of these 
systems with four and eight
remaining supersymmetries in 
four dimensions, respectively,
 are the standard Calabi-Yau
compactifications for which  dilaton 
is constant and the string three-form
field strength vanishes. Recently other
 no-go theorems have appeared 
in the literature for Randall-Sundrum \cite{mc, wz, kl, gl}
and warped compactifications \cite{mn}; 
see also \cite{russo, pope}.

We shall begin our analysis with a summary of the
relevant results of \cite{strom}. 
Then after giving some definitions
of various curvature tensors associated with hermitian
geometry, we shall present our main result. This is an
inequality involving
the length of the torsion of the Chern connection and
the scalar derived by taking twice the trace of $dH$ with the
complex structure of the hermitian manifold. 
We shall then  explore
this inequality in the context of both heterotic
and type II strings.
We shall conclude with some remarks
 regarding the relationship
between the Killing spinor equation 
associated with the
gravitino and that associated with 
the dilatino. In particular,
we shall argue that the former does not always imply the
latter.

The  string compactifications 
we consider  involve supergravity solutions
for which the non-vanishing bosonic fields are those of the
NS$\otimes$NS sector, ie the metric $G$,
 the dilaton $\phi$ and the
string three-form field strength  $H$. In type II strings, 
$H$ is closed but in the
heterotic string due to the anomaly 
cancellation mechanism,
 $dH\not=0$.
In what follows, we shall be mainly 
concerned with compactifications
of the heterotic string though our results can be easily
adapted to the type II strings. 
The relevant heterotic  Killing spinor 
equations in the string frame are
\bea
 \hat\nabla \eta & =& 0
\nonumber\\
\big(\Gamma^M \partial_M \phi & - &{1\over 6}
 \Gamma^{MNR} H_{MNR}\big)\eta = 0\ ,
\label{ks}
\eea
where $M,N,R=0,\dots, 9$, $\Gamma^M$ are the gamma matrices
of spacetime and $\hat\nabla$ is a
 connection with torsion $H$, ie
\beq
\hat\nabla_M Y^N=\nabla_M Y^N+{1\over2} H_{MR}{}^N Y^R\ ;
\eeq
$Y$ is a vector field.
The gamma matrices have been chosen to 
be hermitian along the
space directions and antihermitian along the time direction.
The spinor $\eta$ is in the Majorana-Weyl representation
of the spin group ${\rm Spin}(1,9)$. In the heterotic string,
there is an additional Killing 
spinor equation associated 
with the gaugino. However this will not 
enter in our investigation
and it will be only  very briefly
 mentioned later. In the case of type II
strings there are two additional Killing spinor
 equations to (\ref{ks}) but again
do not enter in our analysis.
It has been shown in \cite{strom} that 
for warped compactifications
 to $\bR^{2(5-n)}$, the solution for the background
 in the string frame can be written as
\bea
ds^2&=&ds^2(\bR^{2(5-n)})+ds^2(M)
\nonumber\\
\phi&=&\phi(y)
\nonumber\\
H&=& {1\over3!}H(y)_{ijk} dy^i\wedge dy^j\wedge dy^k\ ,
\label{ans}
\eea
where $\{y^i; i=1,\dots, 2n\}$ are the coordinates of the
2n-dimensional compact space $M$ and 
$ds^2(M)$ is the metric on $M$.
The manifold $M$ is even-dimensional and 
this will be explained shortly.
The warp factor is not apparent in the 
string frame but in
the Einstein frame because the dilaton 
is not taken to be constant,
the spacetime metric becomes
\beq
ds^2_E= e^{-{1\over2}\phi} ds^2\ ,
\eeq 
where $ds^2_E$ is the  Einstein metric.

For the first Killing spinor equation 
in (\ref{ks}) to have solutions,
the holonomy of the connection $\hat \nabla$ has to be 
an appropriate subgroup of $SO(1,9)$. For the 
compactifications (\ref{ans})
we are considering, the
 non-trivial part of the connection 
$\hat\nabla$ is given by a connection  on the compact
manifold $M$  which we shall 
denote with the same letter.
Therefore for the first Killing 
spinor equation in (\ref{ks})
to have solutions, we take   the  holonomy of $\hat\nabla$  
to be in $SU(n)$. This in 
particular implies
that for compactifications to
 four dimensions $(n=3)$, four 
supersymmetries will be preserved.
The solutions of this Killing spinor equation
are  $\hat\nabla$-parallel
spinors which can be identified with 
the singlets in the decomposition of
the Majorana-Weyl spinor 
representation of $SO(1,9)$ under $SU(n)$.
In addition it was shown in \cite{strom} that $M$ is a 
hermitian manifold, i.e. $M$ admits a  complex 
structure and it is equipped with a 
compatible hermitian
metric. The complex structure is given in terms
of a killing spinor $\eta_+$ as
\beq
J^i{}_j=-i \eta^\dagger_+ \Gamma^i{}_j\eta_+\ ,
\eeq
satisfying
\beq
\hat\nabla_k J^i{}_j=0\ ,
\label{para}
\eeq
i.e. it is parallel with respect to 
the $\hat\nabla$ connection.
The torsion of the connection 
$\hat\nabla$ is determined in terms
of the metric and the complex structure on $M$ as
\beq
H_{ijk}=-3J^m{}_{[i} d\Omega_{|m|jk]}\ ,
\eeq
where $\Omega_{ij}= g_{ik} J^k{}_j$ is the K\"ahler form
of  $M$ and $d\Omega$ is its exterior
derivative, 
$(d\Omega)_{ijk}=3\partial_{[i} \Omega_{jk]}$ and $ds^2(M)=
g_{ij} dy^i dy^j$ is the metric on $M$.
In mathematics $\hat\nabla$ has 
appeared sometime ago  in the
context of hermitian manifolds \cite{yano} 
and recently it is called Bismut
connection \cite{Bi},  and in 
physics has appeared in the 
context of supersymmetric sigma models 
\cite{rocek, hp1, hp2}.
Hermitian manifolds equipped 
with the Bismut connections are also
called KT (K\"ahler with torsion)
 manifolds \cite{hp2}.

Let us now investigate the second
 Killing spinor equation associated
with the dilatino.
For the compactifications we are 
considering, this Killing spinor
equation for $\eta_+$  becomes
\beq
\big(\Gamma^i \partial_i\phi
-{1\over6} \Gamma^{ijk} H_{ijk}\big)\eta_+=0\ .
\eeq
Next we multiply this equation 
from the left and its conjugate
from the right  with $\Gamma^m$, 
respectively. After contracting the two
equations with $\eta_+^\dagger$ 
and $\eta_+$ appropriately and then
substracting them, we 
get schematically
\beq
\eta^\dagger_+\big([\Gamma^m, d\phi]-
{1\over6}\{\Gamma^m, H\}\big)\eta_+=0\ .
\eeq
 After some gamma matrix algebra, this equation gives
\beq
\theta_i=2\partial_i\phi\ ,
\label{thet}
\eeq
where $\theta_i$ is the so called Lee form of $J$ defined
as
\beq
\theta_i=-\nabla^k\Omega_{km} J^m{}_i
={1\over2} J^m{}_i H_{mkn} \Omega^{kn}\ .
\eeq 
We remark that the field and Killing
 spinor equations
of supergravity theory are well-defined 
if we take the dilaton
to be locally defined on $M$ and 
therefore $\theta$ to be closed
but not exact. However in the
 supergravity action,
 $\phi$ appears without a derivative 
 acting on it and therefore
in order the Lagrangian to be a scalar, 
$\phi$ is required to be 
a globally defined scalar on $M$. In many 
backgrounds that arise in the  worldsheet-conformal
field theory approach to
strings, $\phi$ is only locally defined function on the
background. For example, the background associated
with the Wess-Zumino-Witten (WZW) 
model on $S^1\times S^3$ is supersymmetric,
in the supergravity sense, if one introduces a dilaton
which depends linearly on the 
angular coordinate of $S^1$ factor.
Such a dilaton is {\sl not} 
globally defined on $S^1\times S^3$; though
is well-defined in the universal cover of $S^1\times S^3$ 
which is the near horizon geometry of the NS5-brane.
As we shall see in conformal field theory, a more general 
situation can arise.

The key observation in \cite{strom} is that  the existence of
a solution to the supergravity Killing spinor equations  
implies the presence of a holomorphic $(n,0)$-form on $M$.
To see this observe that $M$ 
admits a $\hat\nabla$-parallel $(n,0)$-form
$\epsilon$ because of the requirement that the holonomy
of $\hat\nabla$ is in $SU(n)$.
Next we assume that $\phi$ is a 
{\sl globally-defined} scalar
on $M$ and  write the form
\beq
\tilde \epsilon= e^{-2\phi} \epsilon\ .
\eeq
Then using  $\hat\nabla\epsilon=0$, 
$$
\hat\Gamma^\gamma_{\bar\beta\gamma}=\theta_{\bar\beta}\ ,
$$ 
as it can be verified with an explicit 
computation, and (\ref{thet}), we find
\bea
\partial_{\bar\beta}\tilde\epsilon_{\alpha_1\dots\alpha_n}
& = &- 2\partial_{\bar\beta}\phi
 \tilde\epsilon_{\alpha_1\dots\alpha_n}+
n\hat\Gamma^\gamma_{\bar\beta [\alpha_1}
\tilde\epsilon_{|\gamma|\alpha_2\dots \alpha_n]}
\nonumber\\
&= &- 2\partial_{\bar\beta}\phi
 \tilde\epsilon_{\alpha_1\dots\alpha_n}+
\hat\Gamma^\gamma_{\bar\beta\gamma}
 \tilde\epsilon_{\alpha_1\dots \alpha_n}
\nonumber\\
&= &- 2\partial_{\bar\beta}\phi 
\tilde\epsilon_{\alpha_1\dots\alpha_n}+
\theta_{\bar\beta} \tilde\epsilon_{\alpha_1\dots \alpha_n}=0 \ ,
\eea
where $\alpha_1, \alpha_2, 
\dots \alpha_n, \gamma, \delta=1, \dots, n$
are holomorphic indices with respect to $J$.
So $\tilde \epsilon$ is holomorphic.
This concludes our summary of the 
warped string compactifications.

There are obstructions to the existence of 
holomorphic $(n,0)$-forms on hermitian manifolds.
In particular,  we shall show there does not exist such
a holomorphic (n,0)-form for a large class of the
compactifications we have described.
For this, we first define the Chern connection 
$\tilde\nabla$ on $M$ as
\beq
\tilde\nabla_i Y^j
=\nabla_i Y^j+ {1\over2} J^m{}_i (d\Omega)_{mkn} g^{nj}Y^k\ ,
\eeq
where $Y$ is a vector field of $M$. This
connection is  compatible with the 
metric ($\tilde\nabla g=0$), the complex
structure ($\tilde\nabla J=0$) and the 
holomorphic structure of the
tanget bundle of $M$ (ie the curvature of the 
Chern connection is
a (1,1)-form on $M$). The torsion of the 
Chern connection is
\beq
C_{ijk}={1\over2} 
\big( J^m{}_i d\Omega_{mjk}+ J^m{}_j d\Omega_{imk}\big)\ ,
\eeq
which can be expressed in
 terms of the string three-form $H$ as
\beq
C_{ijk}={1\over4} (J^m{}_i J^n{}_j H_{mnk}-H_{ijk})\ .
\label{candh}
\eeq
Observe that if the torsion of the Chern connection vanishes,
then the manifold $M$ is K\"ahler and consequently $H$ also
vanishes.
Next we define
\beq
b  = -{1\over2}\hat R_{ijkl}\Omega^{ij} \Omega^{kl}\ ,
\eeq
and
\beq
u  = -{1\over4}\tilde R_{ijkl}\Omega^{ij} \Omega^{kl}\ ,
\label{uuu}
\eeq
where $\hat R$ and $\tilde R$ 
are the curvature tensors  of the Bismut and
Chern connections on $M$, respectively.
Moreover a  computation reveals that
\beq
2u=b+C_{ijk} C^{ijk}+{1\over4} dH_{ijkl} \Omega^{ij} \Omega^{kl}\ .
\label{key}
\eeq
(For  the derivation of this see \cite{SIGP}.)
Of course if the holonomy of the
 Bismut connection is contained
in $SU(n)$, then $b=0$.

Necessary conditions on the compact
hermitian manifold $M$ for the existence 
of a holomorphic  $(n,0)$-form, or 
more generally  for the existence of a 
holomorphic sections of a holomorphic 
line bundle over $M$,  
can be derived by the application
of the Gauduchon plurigenera
 theorem \cite{Ga,G4}; for applications
of other vanishing theorems \cite{AI1}  see  \cite{SIGP}. 
These conditions are 
expressed in terms of the Gauduchon metric.
For 
this, we first observe that
given a hermitian manifold $M$ with 
complex structure $J$ and
metric $g$, one can find another 
hermitian structure on $M$
with the same complex structure but 
with metric which is related to $g$
by a conformal rescaling, i.e. metric 
$e^w g$ where $w$ is a function
on $M$. 
It has been shown by Gauduchon \cite{G3}
 that it is possible to always 
find a metric $h$
within the conformal class of $g$ such that 
$$
(\nabla^{(h)})^i\theta^{(h)}_i=0\ ,
$$ 
where  $\nabla^{(h)}$ is the Levi-Civita connection
of  the metric $h$ and $\theta^{(h)}$ is the Lee form
with respect to the metric $h$. 
The metric $h$ that fixes the conformal
gauge is called  {\it Gauduchon metric}. 

The proof for the existence of the 
Gauduchon  metric can be summarized
as follows: Let $(z^1,...,z^n)$ be a holomorphic 
coordinate system on a compact hermitian 
2n-dimensional manifold $(M,J,g)$. 
We consider the complex Laplacian
operator $L_{(g)}$ 
acting on smooth functions $f$ as
 \beq
 L_{(g)}(f)\equiv-2g^{\alpha\bar{\beta}}
\partial_{\alpha}\partial_{\bar{\beta}}f\ .
\eeq
The operator $L$ can be rewritten as
\beq
L_{(g)}(f)=
\Delta f + 2g^{\alpha\bar{\beta}}
\partial_{\alpha}f\theta_{\bar{\beta}}\ ,
\eeq
where $\Delta$ is the standard 
Laplacian associated with the Levi-Civita
connection of $g$ and $\theta$ is the Lee form.
The formal adjoint operator 
$L_{(g)}^*$ of $L_{(g)}$ is given by 
\beq
L_{(g)}^*(f)=
\Delta f - 2g^{\alpha\bar{\beta}}
\partial_{\alpha}f\theta_{\bar{\beta}}- 
2\nabla^{\alpha}\theta_{\alpha}f\ ,
\eeq
where $\nabla$ is the Levi-Civita 
connection of $g$. In particular, we have
\beq
L_{(g)}^*(1)=-2\nabla^{\alpha}\theta_{\alpha}\ .
\eeq
The corresponding operators with
 respect to a conformally equivalent metric
$\bar g = e^{{2\over {n-1}}w} g$ are given by
\bea
L_{(\bar g)}(f)& = & e^{{-2\over {n-1}}w} L_{(g)}(f)\nonumber\\ 
 L^*_{(\bar g)}(f)& = &
e^{{-2n\over {n-1}}w} L_{(g)}^*(e^{2w} f)\ .
\eea
If $L_{(g)}^*(e^{2w})=0$ admits a positive solution, then 
\beq
L_{(\bar g)}^*(1)=-
2\nabla_{(\bar g)}^{\alpha}\theta^{(\bar g)}_{\alpha}=0\ .
\eeq
Thus such $\bar g$ is the Gauduchon metric, $\bar g=h$.

Gauduchon \cite{Ga,G4} showed that  
the equation $L_{(g)}^*(f)=0$ does 
admit a unique positive solution, up to a constant scale,
 by using the following consequence of steps:

 \begin{itemize} 
\item{(i)} If $f$ is in the kernel, 
${\rm Ker} L_{(g)}$, of $L_{(g)}$,  $L_{(g)}(f)=0$, then
$f$ is constant. This follows from 
the Hopf maximal principle.
Roughly if $f$ is not constant, 
then $f$ has a strict maximal point
$p$ on the compact manifold $M$ 
and so $L_{(g)}(f)|_p=\Delta f|_p>0$
 which leads to a contradiction.

\item{(ii)} The index of $L_{(g)}$ vanishes, 
${\rm Index} L_{(g)}={\rm dim Ker}L_{(g)}-
{\rm dim Ker}L^*_{(g)}=0$. This is
because the principal symbols of 
$L_{(g)}$ and of $L_{(g)}^*$ are the same as that of the  
standard Laplacian $\Delta$. Thus using (i), we conclude that
 ${\rm dim Ker} L_{(g)}^*=1$.

\item {(iii)} If a function $f$ is in the image of 
$L_{(g)}$,  $f\in {\rm Im} L_{(g)}$,  then $f$ necessarily changes 
sign as a consequence of the maximum principle. 
Indeed let us suppose that
$f=L_{(g)}k$ for some smooth function $k$ on $M$. 
 Then $k$ has at least a maximum and
at least a minimum point in $M$. If $p$ is a maximum, then 
$f(p)=L_{(g)}(k)|_p=\Delta k|_p>0$.
Now if $p$ is a minimum, then 
$f(p)=L_{(g)}(k)|_p=\Delta k|_p<0$. Therefore $f$
changes sign.  

\item {(iv)} If  $f\in {\rm Ker} L_{(g)}^*$,  
then $f$ is nowhere zero. For this
observe that  ${\rm Ker} L_{(g)}^*$ is orthogonal 
to ${\rm Im} L_{(g)}$. Now suppose that   
 $f$ changes sign, then  one could find $f'$ 
which is either positive or 
negative function of $M$ orthogonal 
 to $f$. Such  function $f'$ then is in ${\rm Im} L_{(g)}$ 
 and so it contradicts (iii).

 \end{itemize}
 
So from (iv) one concludes that there 
is solution of $L_{(g)}^*(f)=0$ which
is a negative or positive function on $M$.
 Observe that if $f$ is negative, then
$-f$ is a positive solution. Next from (ii)
one concludes that the solution 
is unique  up to a constant scale. 
Hence, the Gauduchon metric
 exists and it is unique up to a homothetic 
transformation.

 Now we are ready to state the 
 {\it Gauduchon plurigenera 
theorem} \cite{Ga,G4}. For this let 
us denote with $u^{(h)}$ the scalar $u$ in
  (\ref{uuu})
evaluated for the Gauduchon metric $h$.
If on a compact hermitian manifold 
\beq
\int_M {\sqrt{h}} d^{2n}y \,  u^{(h)}\geq 0\ ,
\label{gaud}
\eeq
then the $mth$-plurigenus 
$p_m(J)=dimH^0(M,{\cal O}(K^m))\in \{0,1\}, m>0$. 
Moreover, if the inequality 
in (\ref{gaud}) is strict, then $p_m(J)=0, m>0$.
 In particular
$h^{n,0}=p_1(J)$ and this gives 
necessary conditions for the existence of 
holomorphic (n,0)-forms.

To briefly explain the proof of the plurigenera theorem, 
let $\lambda$ 
be a holomorphic (n,0)-form. Then we find that 
\beq
0=\partial_{\bar{\beta}}\lambda_{\alpha_1...\alpha_n}=
\tilde{\nabla}_{\bar{\beta}}\eta_{\alpha_1...\alpha_n}
\eeq
 by the properties of 
the Chern connection $\tilde{\nabla}$. 
Next we apply the complex Laplacian $L_{(g)}$ to 
$-{1\over2}|\lambda|^2$, where 
\beq
|\lambda|^2=
\lambda_{\alpha_1...\alpha_n}\lambda^{\alpha_1...\alpha_n}
\eeq
is the square norm of $\lambda$, and get
\bea
L_{(g)}(-{1\over2}|\lambda|^2)& = &
g^{\bar{\beta}\alpha}\partial_{\bar{\beta}}
(\tilde{\nabla}_{\alpha}\lambda_{\alpha_1...\alpha_n}
\lambda^{\alpha_1...\alpha_n})
\nonumber \\
&=&
g^{\bar{\beta}\alpha}\tilde{\nabla}_{\bar{\beta}}
\tilde{\nabla}_{\alpha}\lambda_{\alpha_1...\alpha_n}
\lambda^{\alpha_1...\alpha_n} +
|\tilde{\nabla}\lambda|^2 = 2u|\lambda|^2
 +|\tilde{\nabla}\lambda|^2\ .
\label{kkkk}
\eea
We have used the Ricci identities for the 
Chern connection to derive the 
latter equality.

Now if $u\ge 0$, then $L_{(g)}(-{1\over2}|\lambda|^2)\ge 0$. 
Applying 
the Hopf maximum principle, we find that 
either $\lambda=0$ if $u>0$ or
 $\tilde{\nabla}\lambda=0$ if $u=0$. Hence, in the 
first case $h^{n,0}=0$ and $h^{n,0}\le 1$ in the second.
This observation suffices to show one 
of the main results of this letter
as we shall see below.
However the conditions for the existence of
 holomorphic sections for line bundles
can be put in the form of (\ref{gaud}) as follows: 

Let  $({\cal L},\mu)$ be a holomorphic line bundle
over  a compact hermitian  manifold 
$(M,g,J)$ with fibre metric $\mu$.
We define the  function  $u^{(\mu)}$ 
as  the trace over the $M$ indices 
with respect to $g$ 
of the curvature of the canonical Chern
connection of $({\cal L},\mu)$. For the 
canonical bundle, ie the bundle
of (n,0)-forms on $M$, the function $\mu$ is locally given 
by $\mu={\rm det}(g)={\rm det}(g_{\alpha\bar{\beta}})$. The 
curvature of the canonical bundle is 
\beq
\tilde{R}_{\alpha\bar{\beta}}
=-i\partial_{\alpha}\partial_{\bar{\beta}}{\rm log}({\rm det}(g))
\eeq
 and 
\beq
u^{(\mu)}=-{i\over2} g^{\alpha\bar\beta}
 \tilde R_{\alpha\bar\beta}=-{1\over2}
  g^{\alpha\bar\beta}
  \partial_{\alpha}\partial_{\bar{\beta}}{\rm log}\mu\ .
  \eeq
Observe that if $\mu'=e^w \mu$
 is another fibre bundle
 metric, then $u^{(\mu')}=u^{(\mu)}+L_{(g)}(w)$. 
Suppose that $f_o$ is a 
smooth positive function on  $M$
satisfying $L^*f_o=0$. Gauduchon \cite{Ga} 
showed using the properties of 
the kernel of $L^*$  that the 
average value 
\beq
v_{(g)}({\cal L})={\rm Vol}^{-1}_{(g)}(M)\, 
\int_M\, {\sqrt g}\, d^{2n}y \, f_ou^{(\mu)}
\eeq
 is  independent of the choosen bundle metric $\mu$, 
 where ${\rm Vol}_{(g)}(M)$ is the
  volume of $M$ with respect to the metric $g$. Moreover
  there exist a 
canonical bundle metric $\mu_o$ with  
$u^{(\mu_o)}=v_g({\cal L})$, constant. 
But the existence of holomorphic 
sections of a holomorphic bundle
${\cal L}$ does not depend on the choice of
fibre metric $\mu$, thus it only depends on the sign 
 of the constant $u^{(\mu_o)}$.  
 In the case of interest where
 the line bundle ${\cal L}$ is a 
 positive power of the canonical line 
bundle, 
 the constant 
\beq
 v_g({\cal L})={\rm Vol}^{-1}_{(h)}(M)\,
  \int_M {\sqrt h}\, d^{2n}y 
u^{(h)}\,
\eeq 
where $h$ is the Gauduchon metric. From this one 
can immediately derive the  plurigenera 
theorem as stated in  
(\ref{gaud}).

To apply the above results to string theory, we observe that
if the string metric
 $g$ on $M$ is related to the Gauduchon metric $h$ 
by $g=e^f h$, then we have
\beq
2e^f u=2 u^{(h)}+n(n-1) h^{ij} 
(\theta^{(h)})_i \partial_j f- \Delta^{(h)} f\ ,
\label{akey}
\eeq
where  $\Delta^{(h)}$ is the Laplacian of $h$.
Using (\ref{akey})and (\ref{key}), and
 after integrating by parts,
 we find that
\bea
\int_M {\sqrt{h}} d^{2n}y~~ u^{(h)}
& = &\int_M {\sqrt{h}} d^{2n}y~~  e^f u
\nonumber \\
& =&{1\over2} \int_M {\sqrt{g}} d^{2n}y~~ 
 e^{-(n-1)f}~~\big( C_{ijk} C^{ijk} +{1\over4}
dH_{ijkl} \Omega^{ij} \Omega^{kl}\big)\ .
\eea
This is our main equation. Further, the torsion $C$ of the
Chern connection can be expressed in terms of the string 
three-form $H$ using (\ref{candh}) but the 
above expression will suffice.

Now consider warped compactifications of 
the heterotic string with
compact space the hermitian, but no K\"ahler manifold, $M$ and 
with  spin connection  embedded in the gauge one.
If this is the case, then $dH=0$ and so
\beq
\int_M ~ {\sqrt{h}}~ d^{2n}y~~  u^{(h)}
={1\over2}\int_M~ {\sqrt{g}}~ d^{2n}y~~
 e^{-(n-1)f}~~ C_{ijk} C^{ijk}>0\ .
\eeq
{}From the Gauduchon theorem stated above we conclude
that  $h^{n,0}=0$ and a holomorphic $(n,0)$ form on $M$ cannot
exist. Thus  there are no warped compactifications of the
heterotic string with the spin connection embedded in the
gauge one. 
The same applies for warped  
compactifications of type II strings
for which $H$ is closed, $dH=0$.
{}From these, one concludes 
that the only compactifications involving  the common sector
of type II strings and perturbative 
heterotic string for which the
spin connection is embedded 
in the gauge connection are
those on Calabi-Yau manifolds 
\cite{candelas}. In particular the dilaton
is constant and so there is no warp factor in the metric. 
We remark that the same result can be derived without
using the plurigenera theorem.  
In particular  from (\ref{key}), $dH=0$ and 
the reasoning below equation (\ref{kkkk})
we can conclude that $h^{n,0}=0$ unless $C=0$ and the
manifold is a Calabi-Yau space. 

The only remaining possibility for existence of warped
perturbative  string 
compactifications is that of heterotic
strings for which the spin connection 
is {\it not} embedded in the
gauge connection and so $dH\not=0$. 
A necessary condition 
for having $h^{n,0}=1$
as required is that
\beq
\int_M  {\sqrt{g}} d^{2n}y~~
 e^{-(n-1)f}~~ \big[ C_{ijk} C^{ijk}
 +{1\over4} dH_{ijkl} \Omega^{ij}
\Omega^{kl}\big]\leq 0\ .
\label{ncon}
\eeq
{}From the heterotic string one-loop  
anomaly cancellation formula
\beq
dH=\lambda\big[{\rm Tr} (R'\wedge R')-{\rm Tr} (F\wedge F)\big]\  , 
\eeq
we have
\beq
(dH)_{ijkl} \Omega^{ij} \Omega^{kl}
=-{\lambda\over2}\big({\rm Tr} R'_{ij} R'^{ij}-
{\rm Tr}F_{ij} F^{ij}\big) \ ,
\label{neww}
\eeq
where $\lambda$ is a constant that depends on the string tension,
$R'$ is the curvature of the connection with torsion $-H$,
$F$ is the curvature of a vector bundle on $M$ and the trace
is  taken in  the gauge indices which have been suppressed.
Observe that only the zeroth order term of $H$
in $\lambda$ contributes in the one-loop anomaly and that
$dH=0+{\cal O}(\lambda)$.
To derive (\ref{neww}), 
we have used the fact that 
$\hat R_{ij,kl}=R'_{kl,ij}$ and therefore
 both $R'$ and $F$ are   (1,1)-forms on $M$ which in addition
satisfy  $\Omega^{ij} F_{ij}{}_a{}^b=\Omega^{ij} R'_{ij}{}_k{}^l=0$.
 These conditions on $F$  are derived by solving the Killing
spinor equations associated with the gaugino.
(Global anomaly cancellation requires that $dH$ is exact.)
Now whether or not (\ref{ncon}) is satisfied depends on the details
of the choice of the gauge connection 
and that of the hermitian structure
on $M$.

For  compactifications to six dimensions, 
if $dH=0$ then the only compactifications are those on $K_3$
with constant dilaton \cite{wests}.
On the other hand if $dH\not=0$, then the string metric $g$
is in the same conformal class as the $K_3$ metric. The latter
of course can be identified with the Gauduchon metric.
Therefore $u^{(h)}=0$ and there is a  holomorphic $(2,0)$- form
which is that of $K_3$.

We shall conclude with a remark regarding 
the relationship between
the first and second Killing spinor 
equations in (\ref{ks}). In particular 
we show that the first Killing spinor
 equation does {\it not} always
 imply the second. 
To derive
this, we shall construct an example of a manifold that solves
the first Killing spinor equation but not the second. For this we 
take $M=SU(2)\times SU(2)=S^3\times S^3$; This is a group
manifold model as those in \cite{SSTV, OP1}. Moreover let
as denote the left-invariant one-forms on $M$ with 
$\{(\sigma^r, \bar\sigma^r); r=0,1,2\}$, 
where $\sigma^r$ are associated
with the first $SU(2)$ in $M$ and 
$\bar\sigma^r$ with the second.
In particular, we have
\beq
d\sigma^r=-{1\over2} \epsilon^r{}_{st} \sigma^s\wedge \sigma^t
\eeq
and similarly for $\bar\sigma^r$.
The metric on $M$ can be chosen to be
\beq
ds^2=\delta_{rs} \sigma^r \sigma^s
+\delta_{rs} \bar\sigma^r \bar\sigma^s\ .
\eeq
(There is a two parameter family 
of bi-invariant metrics on $M$ but this
choice will suffice for our purpose.)
The K\"ahler form of the metric that we shall consider is
\beq
\Omega=\sigma^0\wedge \bar\sigma^0
+\sigma^1\wedge \sigma^2+\bar\sigma^1\wedge
\bar\sigma^2
\eeq
In particular we find that the 
string form field strength in this case
is
\beq
H=-\sigma^0\wedge \sigma^1\wedge \sigma^2
-\bar\sigma^0\wedge \bar\sigma^1\wedge\bar\sigma^2\ .
\eeq
The Bismut connection  is the left-invariant connection on
the group $SU(2)\times SU(2)$
and so it has holonomy
 group the identity. Thus its holonomy is
contained in $SU(3)$ and the first Killing spinor equation 
has the required solutions. 
It remains to see whether the 
second Killing spinor equation can
be solved as well. For this we shall attempt to verify the
equation $\theta_i=2\partial_i\phi$.
To do this we compute the Lee form $\theta$ and find
\beq
\theta=\sigma^0-\bar\sigma^0\ .
\eeq
Thus $d\theta\not=0$ and so the second Killing spinor
equation is not satisfied. The manifold $M=SU(2)\times SU(2)$
is associated with a two-dimensional $N=2$ superconformal
theory, for example  that of a WZW model
with target space $M$. Of course such conformal theories do not
have the appropriate central
 charge to serve as string backgrounds,
i.e. the dilaton field equation is  not satisfied.
Nevertheless, it is surprising that 
there is no choice of a dilaton for the above complex
structure, even  after adding for example 
background charges, for which the
associated supergravity background  
preserves some of the spacetime
supersymmetry. 
However if the complex structure
 $J$ is chosen in a different way,
then by adding two appropriate
 background charges, the resulting
background can be thought of as
having geometry $S^1\times S^3\times S^1\times S^3$ which
preserves some spacetime supersymmetry. However 
the dilaton $\phi$ in this case depends linearly 
on the two angular coordinates
of $S^1\times S^1$
and so it is {\it not} a well-defined function
on $S^1\times S^3\times S^1\times S^3$ leading to a $d\phi$
which is a closed but not an 
exact one-form. This is similar to
the case of $S^1\times S^3$ WZW background that has already
been explained. Alternatively, 
one can consider the universal
cover $\bR\times S^3\times \bR\times S^3$ of 
$S^1\times S^3\times S^1\times S^3$.
In such case the dilaton is a 
well-defined function on this background
but the solution is non-compact. All these 
are in accordance with our main result
that there are no  common 
sector type II and heterotic string
warped compactifications 
with the spin connection embedded in the gauge one 
preserving appropriate supersymmetry.

\vskip 0.4truecm
{\bf Acknowledgments:}
We would like to thank G.W. Gibbons and A. Tseytlin 
for helpful discussions. S.I is supported by the contracts 
MM 809/1998 of the Ministry of Science and Education
of Bulgaria, 238/1998 of the University of Sofia ``St. Kl. Ohridski''
and EDGE, Contract HPRN-CT-2000-0010.  G.P.
is supported by a University Research Fellowship from the
Royal Society. This work is partially supported by SPG grant
PPA/G/S/1998/00613.

\end{document}